\documentclass[aps,prd,a4paper,superscriptaddress,amsfonts,
amssymb,amsmath,11pt]{revtex4}
\usepackage{amssymb,latexsym}
\usepackage{amsmath, amsthm}
\usepackage{amscd}
\usepackage{times} 
\usepackage{epsfig} 
\usepackage{psfrag}
\usepackage{graphicx}
\usepackage{mathtools}
\usepackage{pdflscape}
\usepackage[percent]{overpic}

\begin{document}

\title{Black Hole Paradoxes}

\author{Pankaj S. Joshi}
\email{psj@tifr.res.in}

\affiliation{Tata Institute of Fundamental Research, Homi Bhabha Road, 
Colaba, Mumbai 400005, India}
\author{Ramesh Narayan} 
\email{rnarayan@cfa.harvard.edu}
\affiliation{Harvard-Smithsonian Center for Astrophysics, 60 Garden
Street, Cambridge, MA 02138, USA}
\swapnumbers

\begin{abstract}
We propose here that the well-known black hole paradoxes 
such as the information loss and teleological nature of the event horizon are restricted to a particular idealized case, which is the homogeneous dust collapse model. In this case, the event horizon, which defines the boundary of the black hole, forms initially, and the singularity in the interior of the black hole at a later time. We show that, in contrast, 
gravitational collapse from physically more realistic initial conditions typically leads to the scenario in which the event horizon and 
space-time singularity form simultaneously. We point out that this apparently simple modification can mitigate the causality and teleological paradoxes, and also lends support to two recently 
suggested solutions to the information paradox, namely, the `firewall' 
and `classical chaos' proposals.
\end{abstract}

\maketitle

One of the most spectacular predictions of the general theory of
relativity is the black hole, an object that plays a central role in
modern physics [1,2,3]
and astrophysics [4,5].
Black holes are, however,
plagued by fundamental paradoxes that remain unresolved to this day.
First, the black hole event horizon is teleological in
nature [6]
which means that we need to know the entire
future space-time of the universe to determine the current location of
the horizon. This is essentially impossible.  Second, any information
carried by infalling matter is lost once the material falls through
the event horizon. Even though the black hole may later evaporate by
emitting Hawking radiation [7],
the lost information does
not reappear, which has the rather serious and disturbing consequence
that quantum unitarity is violated [8].
Here we propose
that the above paradoxes are restricted to a particular idealized
model of collapse first studied in the 1930s [9, 10]
in
which the event horizon, which defines the boundary of the black hole,
forms initially, and the singularity in the interior of the black hole
forms at a later time. In contrast, gravitational collapse from more
reasonable and/or physically more realistic initial conditions often
leads to models in which the event horizon and the singularity form
simultaneously.  We show that this apparently simple modification
mitigates the causality and teleological paradoxes and at the same
time lends support to two recently proposed solutions to the
information paradox, namely, the ``firewall'' [11]
and ``classical chaos'' [12].

A black hole is expected to form naturally in the universe whenever a
massive star runs out of nuclear fuel at the end of its life and
collapses under its own self-gravity. As a result, the infalling mass
is compressed to a space-time singularity of infinite curvature and
density. This singularity is, however, hidden from view because it is
covered by an event horizon, a one way membrane through which
particles, light rays and signals can enter but from which nothing can
escape.

The above standard picture of black hole formation is motivated by the
classic work of Oppenheimer \& Snyder [9]
and Datt [10],
who studied the collapse of a ``dust'' cloud (fluid
with mass density but no pressure) of uniform density. As the
space-time diagram in Figure~1 shows, the key feature of this
idealized collapse model is that the event horizon forms already at
the space-time point A, where local conditions in the infalling dust
cloud are perfectly normal, and the density and curvature are
finite. The singularity, where the curvature diverges, forms at the
point S, which lies in the future of A; indeed, S lies in the future
for all worldlines within the dust cloud when they cross the horizon,
e.g., the point B. In addition, the location of the horizon at points
A and B depends on the entire future history of infalling matter,
including shells such as C which fall in later and whose very
existence is unknown to the gas falling through A or B. 

This
illustrates the teleological nature of the event horizon: the location
of the event horizon is determined by the entire future history of
space-time, a profoundly paradoxical situation.

\begin{figure}
\begin{center}
\includegraphics[width=0.7\textwidth]{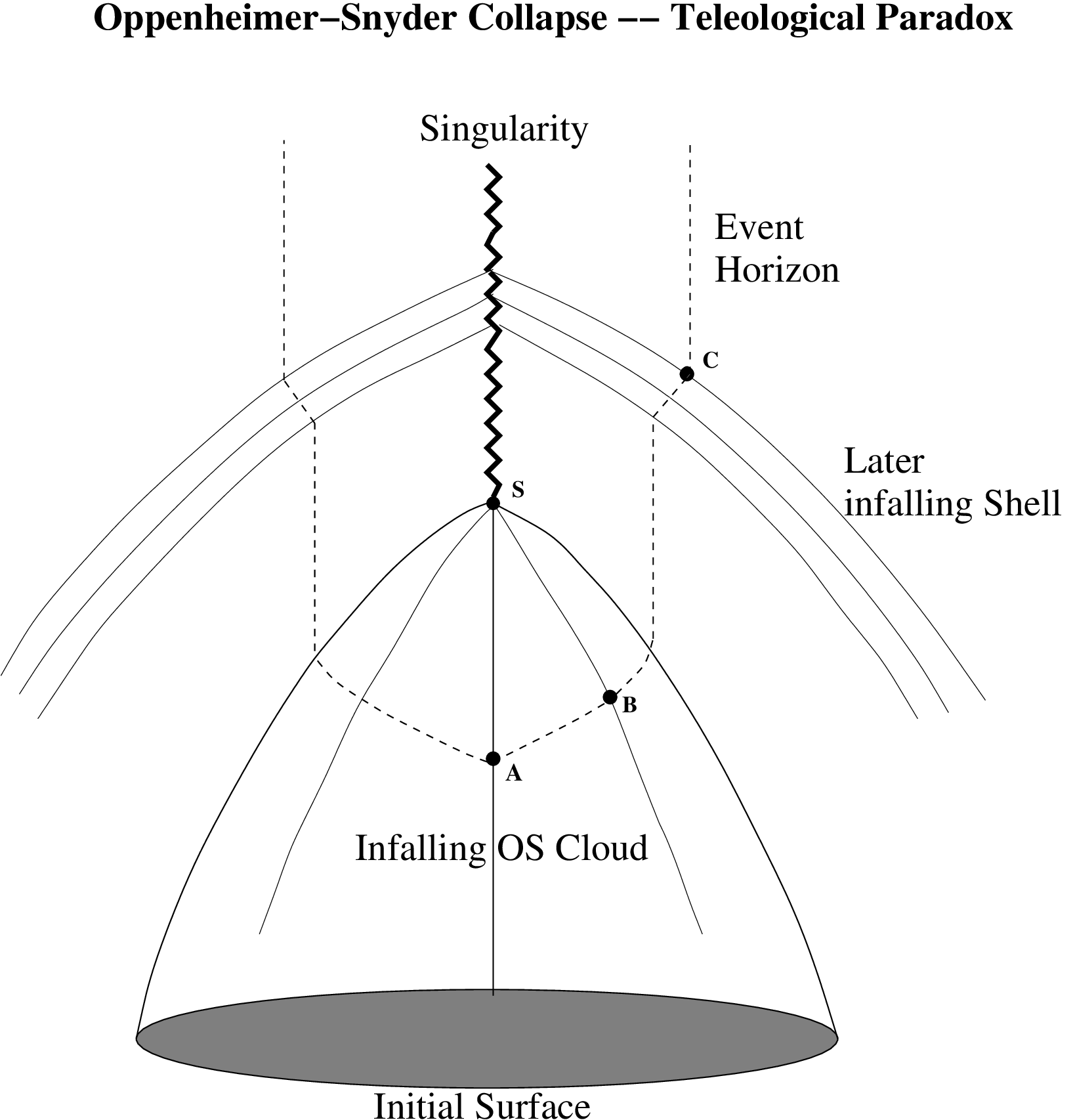}
\caption{Gravitational collapse of a spherical pressureless matter
  cloud (dust cloud) of uniform density whose initial state is shown
  by the shaded region at the bottom. Time increases in the vertical
  direction, and space is represented by the horizontal directions.
  The thin solid lines show worldlines of infalling matter. The event
  horizon (dashed line) first forms on the worldline of the center of
  the cloud, at the space-time point A, where conditions are very
  non-singular. Even point B, on the worldline of matter farther out
  in the cloud, is at an earlier time than the epoch corresponding to
  S, where the singularity (thick jagged line) first forms. Although
  the point C, on a subsequently infalling shell of matter, is at a
  later time than S, no signals can reach this point from the
  singularity. Therefore, if any new physics, e.g. firewall or chaos,
  is to be triggered at A or B or C, it has to be via information
  traveling back from the future, either along the horizon from future
  infinity or along worldlines of matter backward from the
  singularity. This is the teleological paradox, which is caused by
  the fact that the event horizon is causally cutoff from the
  singularity, a consequence of cosmic censorship.}
\end{center}
\end{figure}


Another problem is that black holes run into a major conflict with
quantum theory. A black hole swallows all information carried by
matter falling in through the event horizon. When the hole
subsequently evaporates by emitting Hawking radiation [7],
the mass energy that was swallowed is returned to the external
universe, but in an uncorrelated mixed form that carries no
information.  Thus, a black hole takes in pure quantum states and
converts them to mixed states [8].
This violates quantum
unitarity, which is a very disturbing prospect and is dubbed the black
hole information paradox.

The above paradoxes have attracted considerable attention over the
years and various solutions have been proposed.  A rather radical
solution was suggested recently [11]
in which the event
horizon is replaced by a firewall.  According to this proposal, an
infalling observer encounters a firewall of outgoing bolts of
radiation at the horizon and is destroyed.  Thus, the event horizon is
replaced by a curvature singularity --- the firewall --- associated
with outgoing radiation at the horizon. The information carried in by
the observer is absorbed at the firewall and is presumably returned
via pure quantum states when the singularity radiates or the hole
evaporates, thus solving the information paradox. Proponents of the
firewall hypothesis present it as ``the most conservative
resolution'' [11]
of the information paradox.

The firewall proposal, however, faces several objections, including
the fact that CPT invariance of quantum gravity rules out the
model [12].
In our view, the most serious problem is the
fact that the firewall is by construction located at the event
horizon, but the location of the horizon is determined
teleologically. Somehow, the firewall singularity must sense the
future space-time and thereby decide where it ought to be
located. There is no information, either in local conditions of the
collapsing matter or in signals received from the past, that provides
any indication that a firewall must form.

There are also other models proposed to resolve these issues, 
such as the `fuzzball' scenario (see e.g. 
[13],
and also [14]).
Yet another alternative to the firewall model was suggested 
by Hawking [12]
proposing that gravitational collapse produces only an apparent horizon
and not a true event horizon, and that therefore no information needs
to be lost in the collapse.  In order to avoid the horizon,
Hawking suggests that the region of the collapsed object inside the
event horizon develops a chaotic metric and matter fields.  Such a
chaotic collapsed object would radiate chaotically but
deterministically, so quantum unitarity is preserved. However, the
chaos model again suffers from the teleological problem. Chaos must be
generated at and inside the horizon, but how does the infalling
material know that it should become chaotic when local conditions are
perfectly normal and when the location of the horizon must be
determined by all of the future? Moreover, for near-spherically
symmetric collapse models such as those considered here, any chaos is
likely to be restricted to regions close to the singularity, and the
causal structure of the solution does not permit signals to propagate
from this region out to the horizon.

The problems described above arise as a result of assuming a
foundational principle of black hole theory, namely, the so-called
cosmic censorship conjecture [15].
This conjecture, which is
motivated by the space-time diagram shown in Figure~1, suggests that,
generically, every singularity that forms as a result of collapse is
hidden from view behind an event horizon. That is, the singularity is
always cutoff from the external universe. This has the profound
consequence that it rules out any signals or communication from the
singularity (or its vicinity) to the event horizon. In other words, no
worldline at the horizon ever receives any information from the past
about the singularity; the only way it can find out about the
existence of the singularity is by receiving information from the
future (teleological nature of the horizon).  We propose that this is
the root cause of the black hole paradoxes that we are facing today.

A reasonable alternative emerges if we agree or assume that any
solution to the information paradox, such as firewall or chaos,
requires information to be received at the horizon about the existence
of a singularity in the space-time, and moreover, that this
information should be received via signals from the past. If such
signals are able to reach the event horizon, they could carry
information related to ``new physics'' that might emerge in the
vicinity of the singularity due to the extreme nature of all physical
quantities there, and this information could potentially provide a
causal trigger to generate either a firewall or chaos. By this
reasoning, cosmic censorship and firewalls/chaos are mutually
incompatible, since cosmic censorship requires the future of a
worldline to determine its present behaviour (teleological
communication).

Fortunately, it is now known that the cosmic censorship conjecture is
not fully correct since large classes of physically reasonable
gravitational collapse models have been worked out within Einstein
gravity and been shown to transcend cosmic censorship [16].
In these solutions, the event horizon either is delayed 
or does not form at all [17],
allowing a naked space-time
singularity to be visible to the external universe. Moreover, such
solutions are by no means fine-tuned.  They occur for a wide range of
physically reasonable initial conditions.

To illustrate the last point, we note that the (OSD) collapse shown in
Figure~1 corresponds to an initial density distribution of the dust
cloud of the form
\begin{equation}
\rho(r) = \rho_0, \qquad r \leq r_c.
\end{equation}
This model is physically rather unlikely since it requires a finite
density at the cloud boundary. A homogeneous density sphere is not a
good approximation for a massive star.  Consider instead the following
centrally-peaked model of the initial density, which has the same
radius and mass as the constant density model, but is arguably more
reasonable,
\begin{equation}
\rho(r) = \frac{5}{2}\,\rho_0 \left[1-\left(\frac{r}{r_c}\right)^2
  \right], \qquad r \leq r_c.
\end{equation}
The collapse of this centrally-peaked cloud, starting from dilute
initial conditions ($r_c \gg GM_c/c^2$ where $M_c$ is the total mass
of the cloud), gives the causal structure shown in
Figure~2 [18,19]. 
The key difference from Figure~1
is that the two distinct points A and S in the former have collapsed
to a single point. As a result, the event horizon originates at the
singularity itself, and the singularity becomes naked. Signals emitted
at or near the singularity can have different histories.  An infinite
family of rays escapes to infinity (making the singularity at least
partially visible), some rays are trapped and fall back on the
singularity, and a select few rays travel along the horizon, connecting
the singularity to points on the horizon. The existence of these last
rays --- null geodesics that emerge from the singularity and travel
along the event horizon --- is the key result we highlight here.

\begin{figure}
\begin{center}
\includegraphics[width=0.7\textwidth]{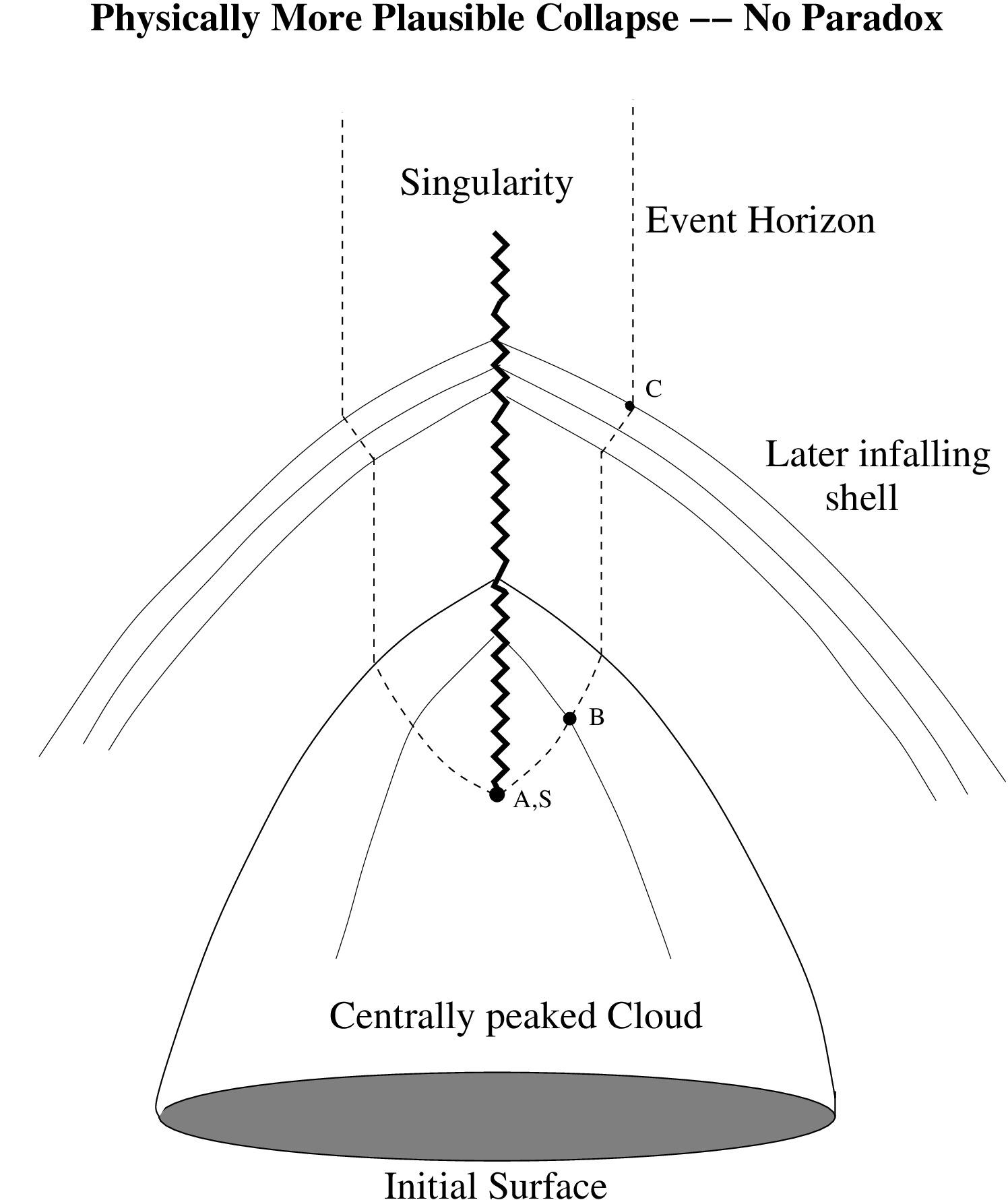}
\caption{Gravitational collapse of a spherical dust cloud with a
  density maximum at the center (equation 2). A naked singularity S
  forms at A, and the event horizon (dashed line) also originates at
  A. As a result, the singularity can communicate with the entire
  event horizon, including points such as B and C. New physics, e.g.,
  firewall or chaos, could thus be triggered on the horizon by signals
  from S, i.e., from the past. Because of the presence of a naked
  singularity, this model violates cosmic censorship, which is rather
  restrictive and is confined to special models such as in Figure
  1. Correspondingly, there is no teleological paradox. A wide range
  of physically reasonable initial conditions of the cloud gives
  collapse with the causal structure shown here.}
\end{center}
\end{figure}

Before exploring the consequences, we note that a wide range of
spherically-symmetric dust collapse models with different initial
density and velocity structure have the same qualitative causal
structure as that shown in Figure~2. Furthermore, models that include
pressure with reasonable equations of state also behave
similarly [20],
as does at least one non-spherically
symmetric model [21].
The general conclusion is that the
causal structure shown in Figure~2 is as plausible as the standard OSD
solution (Figure~1), and possibly more realizable in realistic
physical situations.  One caveat is that models with rotation have not
so far been explored, so their collapse properties are not known.

The model illustrated in Figure~2 produces a radically different
scenario compared to the OSD picture of collapse (Figure~1) which has
formed the basis of all discussions so far on the causal structure of
black holes, the information paradox, cosmic censorship. etc.  Fluid
at the center of the cloud (initial density peak) no longer enters the
horizon when it is physically very regular with modest density and
curvature. Instead, by the time this matter reaches the horizon, it
has already attained extremely high densities and has an arbitrarily
large space-time curvature. We expect the matter to behave very much
like the hot big bang in reverse, and to become arbitrarily hot and
radiation-dominated. Moreover, as the curvature approaches the Planck
scale (or other appropriate scale), new physical phenomena associated
with quantum gravity should emerge. Most importantly, signals from
this ultra-dense region in the quantum gravity regime will flow out
along the horizon, conceivably modifying physics throughout the
horizon.  Two ways in which this might happen are illustrated in
Figure~3.


\begin{figure}
\begin{center}
\includegraphics[width=0.7\textwidth]{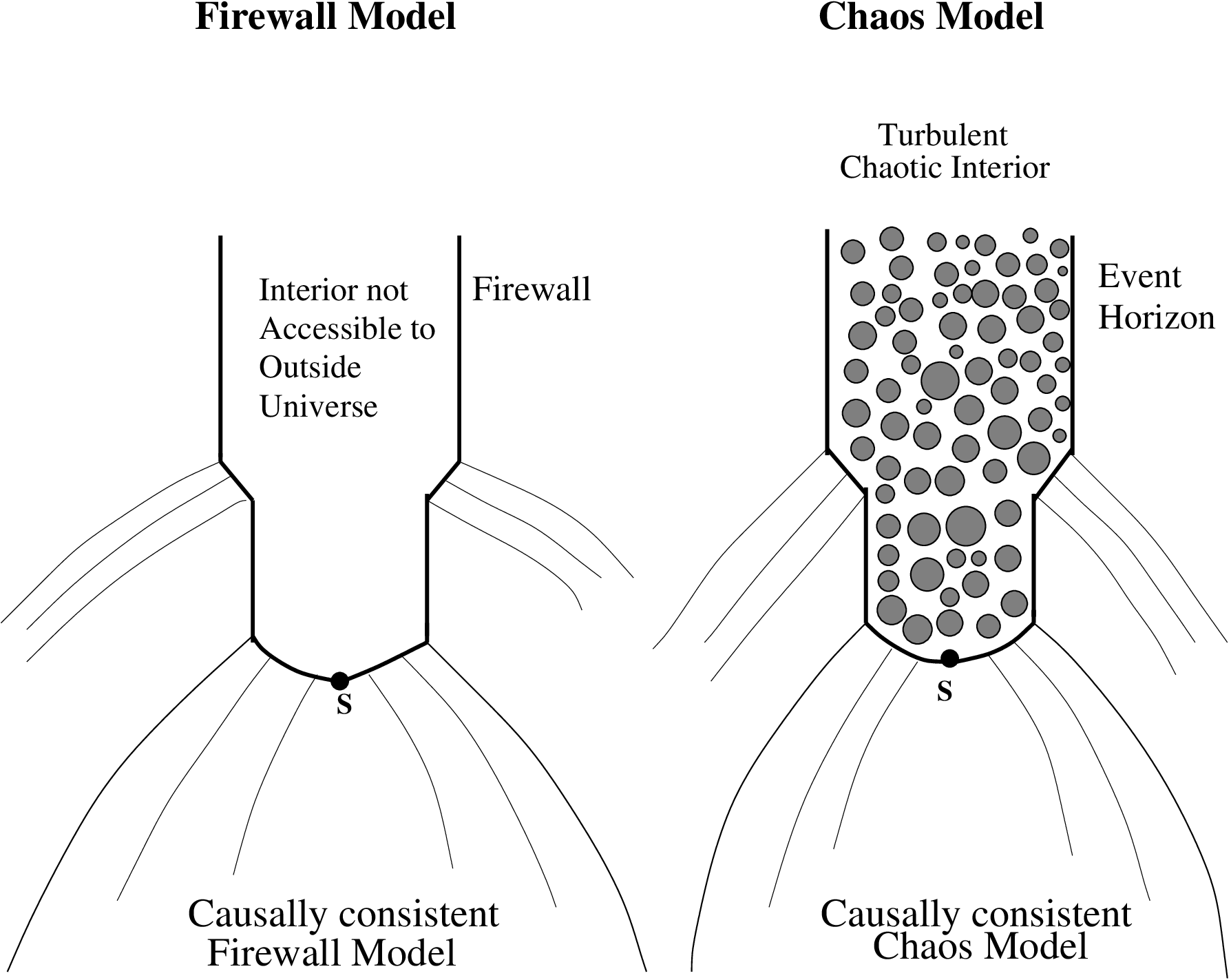}
\caption{A causally consistent version of the firewall (left) and
  chaos (right) models. Both models require some exotic phenomenon ---
  a singular firewall or chaotic dynamics --- to switch on suddenly at
  the horizon (thick solid lines). To trigger this behaviour, a
  warning of some sort must reach fluid that is about to cross the
  horizon. Because of cosmic censorship, no such warning from the past
  is possible if the collapse behaves as in Figure 1. In contrast,
  when the event horizon connects to the singularity in the past, as
  in the case of the naked singularity scenario shown in Figure 2,
  signals from the singularity travel to all points on the horizon.
  Since the singularity is a region of extreme physical conditions
  corresponding to the quantum gravity regime, signals originating
  from here could, in principle, trigger the necessary behaviour for
  the firewall and chaos models. (No specific trigger mechanism is
  described here since the focus is on establishing causality.)}
\end{center}
\end{figure}

In one scenario, the quantum matter at the singularity in Figure~2 is
radiated away along outgoing rays. The maximum burst of radiation will
arguably be along the event horizon because close to that surface and
below the singularity the densities, pressures and all other physical
quantities attain their maximum and unbounded values. The firewall
could then originate at the naked singularity and propagate as a
singular wall of outgoing radiation. As material farther out in the
cloud approaches the firewall, e.g., point B, even though its own
local properties may be quite regular, when it hits the singularity at
the firewall, it will be absorbed and will add to the firewall
(Figure~3 left). We do not have any explanation of how the latter
might happen, but neither was an explanation offered with the original
firewall hypothesis. Our contribution here is to show that it is
possible to have a firewall originate at a singular point and then
evolve causally, without any need for a teleological connection to the
future.

Another interesting point is that the null and timelike paths in the
vicinity of the naked singularity have been examined in earlier
work [22,23]
and found to have a complex
structure that might well give rise to chaotic behaviour in the
interior of the horizon. If this chaos, which will originate in the
naked singularity, is able to expand and fill the interior of the
horizon, then it is conceivable that the entire interior of the black
hole could become chaotic, thereby resembling the classical chaos
model [12]
(Figure~3 right). Once again, we do not attempt
to explain the detailed dynamics of such a model. What we show is that
a chaos model could potentially be causally consistent without having
to invoke teleological properties.

With reference to chaos from gravitational collapse, we know that the
inner region of Kerr geometry is unstable and might be chaotic.  If
the collapse of a massive rotating star results in a configuration
described by the Kerr metric on the outside, this may offer a way to
produce chaos in the interior, as suggested by
Hawking [12].
However, we note that the unstable region
does not extend outside the inner horizon and so chaos cannot
propagate all the way to the outer horizon as needed to solve the
information paradox. Thus the chaos model requires some other trigger
to generate the necessary turbulence. Our proposal in Figure 3 is a
possible solution.

In summary, we suggest that some of the problems that have plagued
black hole physics might be the result of (i) relying on the classical
OSD picture of gravitational collapse of a constant density cloud
(Figure~1), in which the event horizon forms much earlier than the
singularity, and (ii) assuming that this model and its associated
cosmic censorship describes the generic behaviour of collapse. By
making use of physically more realistic gravitational collapse models,
e.g., those with initial density higher at the center (Figure~2), a
very different picture of black hole formation through collapse
emerges in which the horizon and the singularity generically form at
the same epoch. Such models violate cosmic censorship and can
potentially resolve the event horizon and information paradoxes.

Finally, a number of astronomical observations indicate that
astrophysical black holes have dark surfaces, and this has been used
to argue that they possess event
horizons [4,24,25].
In actual fact, the
evidence indicates only that the objects possess apparent horizons, so
the observations are in principle consistent with the chaos
model [12].
The situation is less clear in the case of the
firewall model [11],
since there has been little
discussion of the radiative properties of the firewall.
\bigskip

Acknowledgment: RN thanks the Tata Institute of Fundamental Research for
   hospitality while this work was completed, and the NSF for partial
   support under grant AST1312651.

\bigskip 


REFERENCES
\bigskip


[1] Susskind, L. and Lindesay, J.,
    "An introduction to black holes, information, and the string 
theory revolution : the holographic universe", World Scientific, (2005).


[2] Page, D. N., "Hawking radiation and black hole 
thermodynamics", New Journal of Physics, (2005), {\bf 7}, p.203.


[3] Sen, A., "Black hole entropy function, 
attractors and precision counting of microstates", GRG Journal,
(2008), {\bf 40}, p. 2249.


[4] Narayan, R., "Black holes in astrophysics",
  New Journal of Physics, (2005), {\bf 7}, p199.


[5] Begelman, M. and Rees, M.,
    "Gravity's Fatal Attraction", (2010),
Cambridge, UK: Cambridge University Press.


[6] Hawking, S. W. and Ellis, G. F. R.,
    "The large-scale structure of space-time.",
Cambridge (UK): (1973), Cambridge University Press.


[7] Hawking, S. W., "Black hole explosions?", Nature,
   (1974), {\bf 248}, p30.


[8] Hawking, S. W., 
``Breakdown of predictability in gravitational collapse", Phys. Rev.D,
 (1976), {\bf 14}, p 2460.


[9] Oppenheimer, J. R. and Snyder, H.,
   "On Continued Gravitational Contraction",
  Phys. Rev. (1939), {\bf 56}, p455.


[10] Datt, B., "Uber eine Klasse von L{\"o}sungen 
der Gravitationsgleichungen der Relativit{\"a}t", Zeitschrift fur Physik,
   (1938), {\bf 108}, p314.


[11] Almheiri, A. and Marolf, D. and Polchinski, J. 
and Sully, J., "Black holes: complementarity or firewalls?",
  Journal of High Energy Physics, {\bf 2}, p62.


[12] Hawking, S. W.,
   "Information Preservation and Weather Forecasting for Black Holes",
  ArXiv e-prints 1401.5761,


[13] Mathur, S., see e.g. http://arXiv.org/abs/arXiv:1308.2785 
and references therein.

[14] Kalyana Rama S., e-Print: arXiv:1211.5645 [hep-th]



[15] Penrose, R.,
    "Gravitational Collapse: the Role of General Relativity",
  Nuovo Cimento Rivista Serie, (1969), {\bf 1},
    p252.


[16] Joshi, P. S.,
    "Gravitational Collapse and Spacetime Singularities",
Cambridge University Press, Cambridge, UK, 2007.


[17] Joshi, P. S. and Malafarina, D. and Narayan, R.,
    "Distinguishing black holes from naked singularities through 
their accretion disc properties",
  Classical and Quantum Gravity, (2014), {\bf 31},
       p015002.


[18] Joshi, P. S. and Dwivedi, I. H.,
    "Naked singularities in spherically symmetric inhomogeneous 
Tolman-Bondi dust cloud collapse",
  Phys. Rev. D, {1993), {\bf 47}, p5357.


[19] Deshingkar, S. S. and Jhingan, S. and 
Joshi, P. S., "On the Global Visibility of the Singularity in 
Quasi-Spherical Collapse", GRG Journal (1998), {\bf 30},
    p1477.


[20] Joshi, P. S. and Malafarina, D.,
    "Recent Developments in Gravitational Collapse and 
Spacetime Singularities", International Journal of Modern Physics D,
(2011), {\bf 20},
    p2641.


[21] Joshi, P. S. and Kr{\'o}lak, A.,
    "Naked strong curvature singularities in Szekeres spacetimes",
  Classical and Quantum Gravity, (1996), {\bf 13},
    p3069.


[22] Deshingkar, S. S. and Joshi, P. S.,
``Structure of nonspacelike geodesics in dust collapse",
Phys. Rev. D, (2001), {\bf 63}, p.024007.


[23] Deshingkar, S. S. and Joshi, P. S. 
and Dwivedi, I. H.,
 "Appearance of the central singularity in spherical collapse",
  Phys. Rev. D, (2002), {\bf 65}, 
p084009.


[24] Narayan, R. and McClintock, J.~E.,
    "Advection-dominated accretion and the black hole event horizon",
  New Astron. Rev., (2008), {\bf 51},
    p733.


[25] Broderick, A.~E. and Loeb, A. and Narayan, R.,
    "The Event Horizon of Sagittarius A*", Astrophys. J., 2009, {\bf 701},
    p 1357-1366.

\end{document}